\documentclass[usenatbib]{raa}            
\usepackage{graphicx,times}
\usepackage{natbib}
\usepackage{hyperref}
\newcommand{\ergs}{\ifmmode {\rm erg\ s}^{-1} \else erg s$^{-1}$\ \fi}

\newcommand{\lb}{\ifmmode L_{\rm Bol} \else $L_{\rm Bol}$\ \fi}
\newcommand{\ledd}{\ifmmode L_{\rm Edd} \else $L_{\rm Edd}$\ \fi}
\newcommand{\leddR}{\ifmmode L_{\rm Bol}/L_{\rm Edd} \else $L_{\rm Bol}/L_{\rm Edd}$\ \fi}
\newcommand{\lx}{\ifmmode L_{\rm 2-10keV} \else  $L_{\rm 2-10keV}$\ \fi}
\newcommand{\hb}{\ifmmode H\beta \else H$\beta$\ \fi}
\newcommand{\ha}{\ifmmode H\alpha \else H$\alpha$\ \fi}
\newcommand{\hg}{\ifmmode H\alpha \else H$\gamma$\ \fi}
\newcommand{\oiii}{[O {\sc iii}]\ }
\newcommand{\oii}{[O {\sc ii}]\ }
\newcommand{\nii}{[N {\sc ii}]\ }
\newcommand{\sii}{[S {\sc ii}]\ }
\newcommand{\neiii}{[Ne {\sc iii}]\ }
\newcommand{\mbh}{\ifmmode M_{\rm BH}  \else $M_{\rm BH}$\ \fi}
\newcommand{\lv}{\ifmmode \lambda L_{\lambda}(1350\AA) \else $\lambda L_{\lambda}(1350\AA)$\ \fi}
\newcommand{\lcon}{\ifmmode L_{1350} \else $L_{1350}$\ \fi}

\newcommand{\mdot}{\ifmmode \dot{m} \else \dot{m} \fi }
\newcommand{\llog}{\ifmmode {\rm log} \else {\rm log} \fi }
\newcommand{\kms}{\ifmmode {\rm km\ s}^{-1} \else km s$^{-1}$\ \fi}

\newcommand\HST{{\it{HST\ }}}

\begin{document}

   \title{The inner $\sim$ 40 pc Radial Distribution of the Star formation
Rate for a nearby Seyfert 2 galaxy M51
}

 \volnopage{ {\bf 20xx} Vol.\ {\bf 9} No. {\bf XX}, 000--000}
   \setcounter{page}{1}

   \author{Li-Ling Fang
   \and Xiao-Lei Jiang
   \and Zhi-Cheng He
   \and Wei-Hao Bian
      \inst{1}
   }

   \institute{Department of Physics and Institute of Theoretical Physics, Nanjing
Normal University, Nanjing 210046, China; {\it whbian@njnu.edu.cn}\\
\vs \no
   {\small Received [year] [month] [day]; accepted [year] [month] [day] }
}

\abstract{We investigate spatially resolved specific star formation
rate (SSFR) in the inner $\sim$ 40 pc for a nearby Seyfert 2 galaxy,
M51 (NGC 5194) by analyzing spectra obtained with the \emph{Hubble Space
Telescope (HST)} Space Telescope Imaging Spectrograph (STIS). We present 24
radial spectra measured along the STIS long slit in M51,
extending $\sim 1\arcsec$ from the nucleus (i.e., -41.5 pc to 39.4
pc). By the simple stellar population synthesis, the stellar
contributions in these radial optical spectra are modeled. Excluding
some regions with zero young flux fraction near the center
(from -6 pc to 2 pc), we find that the mean flux fraction of young stellar populations
(younger than 24.5 Myr) is about 9 \%, the mean
mass fraction is about 0.09\%. The young stellar populations are not
required in the center inner $\sim$ 8 pc in M51, suggesting a
possible SSFR suppression in the circumnuclear region ($\sim$ 10
pc) from the feedback of active galactic
nuclei (AGNs). The radial distribution of SSFR in M51
is not symmetrical with respect to the long slit in STIS. This
unsymmetrical SSFR distribution is possibly due to the unsymmetrical
AGN feedback in M51, which is related to its jet.
 }

   \authorrunning{Fang, Jiang, He \& Bian}            
   \titlerunning{Radial SSFR in M51}  
   \maketitle


%
%
\section{INTRODUCTION}
The star formation history in galaxies is intimately related to the
galaxy evolution, as well to the evolution of central super massive
black holes \citep[e.g.][]{Kennicutt98, Tremaine02, Asari07, Chen09}. It has been suggested that star
formation in circumnuclear regions in Seyfert galaxies is
suppressed \citep[e.g.][]{Wang07}. Also, specialists in this field have found that there is a
correlation between the mean specific star formation rate and the
Eddington ratio, suggesting that active galactic nuclei (AGNs) are
possibly triggered by supernova explosions in the circumnuclear
regions \citep[e.g.][]{Chen09}.

With high angular resolution, radial spectra obtained with \emph{Hubble
Space Telescope (HST)} Space Telescope Imaging Spectrograph (STIS)
provide unique information on the host and central accretion processes
in active galaxies (e.g. Spinelli et al. 2006; Rice et al. 2006).
For some nearby Seyfert 2 galaxies, their central engines are
obscured by a torus, which enhances the contrast between the
stellar light from their host and the nuclear continuum from the
central accretion disk. In order to investigate radial star
formation rate in circumnuclear regions of AGNs, a nearby Seyfert 2
galaxy, M51, is selected, and spectra from its central region ($\sim$ 40 pc)
obtained from \HST STIS are analyzed.

The paper is organized as follows: \S 2 presents the data reduction
and the radial spectra. \S 3 models the spectra by simple
stellar population synthesis (SSPS). \S 4 includes our discussion and
results.

\section{Data Reduction}

\begin{table*}
\centering \caption{M51 observations by HST STIS.}
\begin{tabular}{lrrrrrrrr} \hline \hline
Data & dataset & Exposure Time & Grating & Aperture & Center Wavelength & $\Delta \lambda$\\
\hline
1998 Apr 2 & O4R401090 & 1858 s & G430M & $52"\times 0."2$ & $4961 \AA$ & 282\AA \\
1998 Apr 2 & O4R4010A0 &  600 s & G430M & $52"\times 0."2$ & $4961 \AA$ & 282\AA \\
1998 Apr 2 & O4R4010B0 & 2066 s & G430L & $52"\times 0."2$ & $4300 \AA$ & 2807\AA \\
1998 Apr 2 & O4R4010C0 & 1020 s & G430L & $52"\times 0."2$ & $4300 \AA$ & 2807\AA \\
1998 Apr 2 & O4R4010D0 & 1644 s & G750L & $52"\times 0."2$ & $7751 \AA$ & 4987\AA \\
1998 Apr 2 & O4R4010E0 &   25 s & G750L & $52"\times 0."2$ & $7751 \AA$ & 4987\AA \\
\hline
\end{tabular}
\end{table*}

\subsection{\emph{HST} STIS data}
M51 (NGC 5194) is a nearby Seyfert 2 galaxy with z=0.0015 (e.g.,
Spinelli et al. 2006). Searching the \HST archive for M51, there are
six datasets observed by STIS, i.e., two by G430M, two by G430L, two by
G750L (Table 1). Checking these data, we find that two datasets
(i.e., O4R4010B0 by G430L and O4R4010D0 by G750L) acquired in low dispersion
grating modes are suitable for host stellar population analysis,
which provides continuous wavelength coverage spanning the spectral
region from 2900~\AA\ to 1$\mu m$ \citep[e.g.][]{Bradley04}. The dataset
by G430M has very narrow wavelength coverage ($\sim$4800\AA -
5100\AA), which is not suitable for SSPS. The STIS plate scale is
$0.05078\arcsec \rm pixel^{-1}$ and the slit length is 50\arcsec.
Given the distance to M51 of 8.4 Mpc, 1\arcsec corresponds to 40.7
pc and 1 pixel corresponds to 2.07 pc.

These STIS datasets by G430L and G750L give high spatial
resolution (0.1\arcsec, or 4.1~pc) spectra in the central region of M51.
The spectra were acquired through a $52\arcsec \times 0.\arcsec 2$
orientation at a single position angle of 166\dg, and the radio
structure is elongated along a position angle of 166\dg\ (east of
north, Bradley et al. 2004). The 0.2\arcsec wide slit was projected onto
3.75 STIS CCD pixels, which corresponds to spectral resolutions of
10.3~\AA\ for G430L and 18.3~\AA\ for G750L \citep[e.g.][]{Bradley04}.

Because the active nucleus of M51 is obscured, which is expected for
a Seyfert 2 galaxy, the AGN continuum is not apparent in our lower
resolution G430L and G750L spectra. Instead, the circumnuclear spectra
exhibit strong stellar population in the M51 host galaxy.

\subsection{Data reduction in IRAF}

From STIS website, calibrated files were downloaded, as well as the
reference files. Following STIS Data Handbook, one-dimensional spectra
were extracted from the two-dimensional spectra using the CALSTIS {\tt x1d}
routine, which performs a geometric rectification and background
subtraction. In order to investigate the radial spectra at different
distances from the nucleus, we extracted spectra at different
positions and with different sizes of the extraction box, in
reference pixels. Following Bradley et al. (2004), the nucleus of
M51 was selected between cloud 4 and cloud 5 (these clouds
correspond to the narrow line regions in M51).

For STIS grating modes G430L and G750L, we extracted spectra at
the same positions with the same sizes as the extraction boxes. For
G430L spectra, the wavelength coverage was 2901 \AA -- 5709 \AA, and
for G750L spectra, it was 5267 \AA -- 10258 \AA. Considering the
balance between the number of radial spectra and the spectral
signal-to-noise ratio (snr), the extraction box was from 1 to 4
pixels, corresponding to angular sizes of $0.051'' - 0.204''$ (line
3 in Table 2). There were 24 spectra extracted at both sides of the
nucleus, extending $\sim$ 1\arcsec (40.7 pc) from the nucleus (Fig.
1, 2).

\subsection{Simple Stellar Population Model}

\begin{figure*}
\begin{center}
\includegraphics[width=12cm,angle=-90]{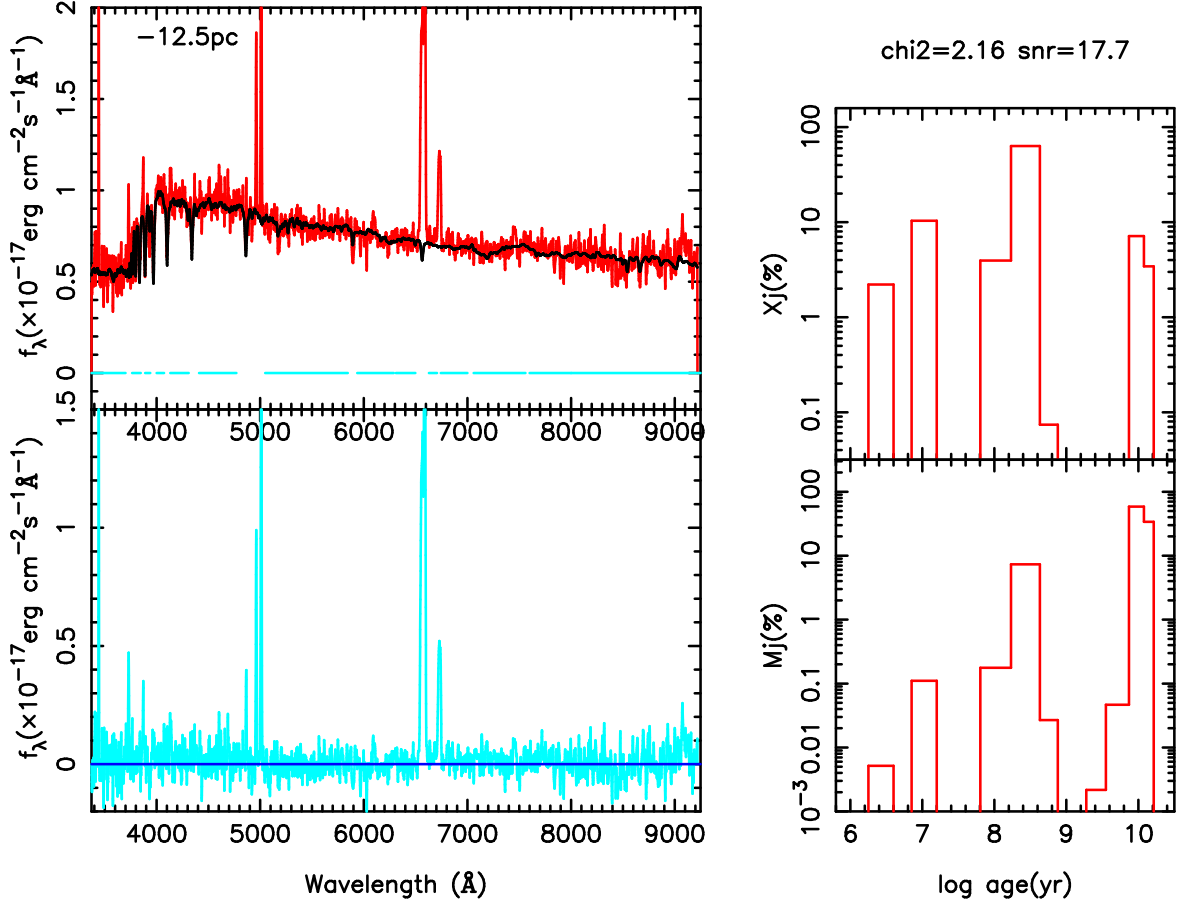}
\caption{An example of the simple stellar population synthesis for a
spectrum separated from the nucleus by -12.5 pc. Top left: Observed (the
red line) and model (the black line) spectra. Cyan line in the
bottom indicates the fit windows. Bottom-left: Residual spectrum.
Right: the distributions of flux fraction (top) and mass fraction (bottom) as a
function of stellar age. The reduced $\chi^2$ and snr are given at the top of the right
panel.}
\end{center}
\end{figure*}

Spectra acquired by G430L and G750L have different wavelength coverages. However, there is an overlap between the spectra acquired by G430L and G750L. We used the
following step to process data from Simple Stellar Population (SSP) model. First, all spectra were resampled with $\Delta \lambda=1$ \AA. Second, the
spectra acquired by G750L were scaled to have the same mean flux as those by
G430L in their common wavelength range of 5500 \AA\ -- 5600 \AA. For
this common wavelength range, we used the spectra by G430L. Then we
obtained 24 radial spectra from M51, with a wavelength coverage of 2901
\AA -- 10258 \AA\ (Fig. 1 and 2).

We use SSP synthesis \citep[STARLIGHT;][]{Cid05} to model
the stellar contribution in the Galactic extinction-corrected
spectrum in the rest frame \citep{Cid05, Bian10}. The Galactic extinction law of \cite{Cardelli89} 
with $R_{V} = 3.1$ is adopted. There are three files used by
the code STARLIGHT, i.e., a base-file, a configuration file and a
mask-file for emission lines. The configuration file contains all of
the technical parameters which control STARLIGHT, including the clipping
threshold, limits for extinction and kinematical parameters, and limits
for the intrinsic extinction.

We use 45 default templates in \cite{Cid05}, which
are calculated from the model of \cite{Bruzual03} with a
spectral interval of 1\AA\ between 3322\AA\ to 9300\AA. The linear
combination of 45 templates is used to represent the host bulge
spectrum. These 45 templates are comprised of 15 ages, from 1 Myr to 13 Gyr,
i.e., $t=$ 0.001, 0.00316, 0.00501, 0.01, 0.02512, 0.04, 0.10152,
0.28612, 0.64054, 0.90479, 1.434, 2.5, 5, 11 and 13 Gyr, and three
metallicities, $Z=$ 0.2, 1 and 2.5 $Z_\odot$ (Cid Fernandes et al.
2005). At the same time as the SSP fit, we add a power-law component
with a fixed slope of -0.5 ($f_{\lambda} \propto \lambda^{-0.5}$) in
the code to represent the AGN continuum emission. Information about these 45
templates is listed in the base-file, which is used in STARLIGHT.

The synthetic spectrum is built using the following equation,
\begin{equation}
M_\lambda = M_{\lambda_0}
   (\sum_{j=1}^{N_*} x_{\rm j} b_{\rm j,\lambda} r_\lambda)
   \otimes G(v_0,v_d)
\end{equation}
where $b_{\rm j,\lambda}$ is the $j^{\rm th}$ template normalized at
$\lambda_0=4020$ \AA, $x_j$ is the flux fraction at 4020 \AA,
$M_{\lambda_0}$ is the synthetic flux 4020 \AA, $r_{\lambda}\equiv
10^{-0.4(A_{\lambda}-A_{\lambda 0})}$ is the reddening term by
V-band extinction $A_V$ adopted for the Galactic extinction law, and
$G(v_0,v_d)$ is the line-of-sight stellar velocity distribution,
modeled as a Gaussian centered at velocity $v_0$ and broadened by
the velocity dispersion $v_d$. The limit is from 0 to 500
\kms for $v_d$ , from -500 to 500 \kms for $v_0$, from 0 to 5 mag for $A_V$,
which are listed in the configuration file.

We exclude the AGN emission lines in the mask-file, such as H Balmer
lines, \oii$\lambda$3727, \neiii$\lambda$3869, \oiii$\lambda
\lambda$4959, 5007, \nii$\lambda \lambda$6548, 6583, and \sii$\lambda
\lambda$6717, 6731.  We also limit rest-frame wavelength range from
3372\AA\ to 9000\AA\ to match the SSP templates.

The best fit is reached by minimizing reduced $\chi^2$ between the
observed spectrum and the model spectrum with a simulated annealing
plus Metropolis scheme. An example fit is shown in Fig. 1.

\section{Results and Discussion}

\begin{table*}
\centering \caption{The properties of 24 radial spectra in M51. For
the last four lines, young\_x1 and young\_m1 are respectively the flux fraction and the mass fraction for $t < 0.102~Gyr$, young\_x2 and young\_m2 are respectively the flux fraction and the mass fraction for $t < 24.5~Myr$. }
\begin{tabular}{lccccccccccccccccccc} \hline \hline
Num&(1)&(2)&(3)&(4)&(5)&(6)&(7)&(8)&(9)&(10)&(11)&(12)\\ \hline
Dis(pc)&-41.5&-37.4&-33.2&-29.1&-24.9&-22.8&-20.8&-18.7&-16.6&-14.5&-12.5&-8.3\\
Aper(pix)&2&2&2&2&1&1&1&1&1&1&1&2\\
Snr&6.7&7.8&7.4&8.4&14&15.3&11.4&13.8&10.1&14.3&17.7&8.8\\
$\chi^2$&2.51&2.31&0.78&0.87&2.80&3.20&1.32&1.58&0.85&1.47&2.16&0.34\\
$A_v$ & 1.23 & 1.27 & 0.91 & 1.32& 0.78& 0.91 &0.79
&0.69&0.66&0.53&0.39&0.37\\
pl\_x\%&0.00&0.00&0.00&0.00&0.00&0.02&6.74&2.60&5.10&4.17&10.44&15.70\\
young\_x1 (\%) & 0.22&  0.00& 15.95& 15.74& 65.24& 80.52& 80.81&
58.67& 46.20& 42.90& 16.55&  8.97\\
young\_x2 (\%) & 0.22&  0.00& 10.56& 13.48&  1.30&  7.80&  3.51&
15.69& 24.14& 13.83& 12.56&  8.97\\
young\_m1 (\%) &0.001&0.00&0.257&1.042&2.716&4.501&5.407&2.247&1.496&1.540&0.294&0.048\\
young\_m2 (\%)&0.001&0.00&0.085&0.627&0.011&0.088&0.053&0.114&0.087&0.161&0.116&0.048\\
\hline \hline
Num&(13)&(14)&(15)&(16)&(17)&(18)&(19)&(20)&(21)&(22)&(23)&(24)\\
\hline
Dis(pc)&-4.1&center&4.1&8.3&12.5&14.5&16.6&18.7&20.8&24.9&31.3&39.4\\
Aper(pix)&2&2&2&2&1&1&1&1&1&2&4&4&\\
Snr&6.6&10.4&12.9&11.1&12.2&12.3&10&16.7&10.4&8.2&6.74&8\\
$\chi^2$&0.20&0.71&1.52&0.57&1.24&0.92&0.59&1.87&0.83&0.36&0.35&1.64\\
$A_v$ &0.41 &0.82
&1.45&1.21&1.01&0.74&0.70&0.37&0.53&0.76&1.27&1.46\\
pl\_x\%&21.58&15.09&19.88&14.94&0.00&7.45&1.42&0.00&3.74&3.84&0.0&0.00\\
young\_x1 (\%) & 0.08& 23.30& 31.58& 52.12& 19.34& 16.31& 54.15& 25.20& 13.93& 62.38& 25.35&  0.00\\
young\_x2 (\%) & 0.00&  0.00&  4.14& 11.72& 19.30& 16.07& 14.13& 11.88& 13.93&  0.01&  7.71&  0.00\\
young\_m1 (\%) &0.000&0.588&1.042&1.832&0.101&0.151&2.107&0.456&0.093&2.139&0.802&0.000\\
young\_m2 (\%) &0.000&0.000&0.038&0.112&0.100&0.140&0.04&0.08&0.09&0.000&0.080&0.000\\
\hline
\end{tabular}\\

\end{table*}

\begin{figure*}
\begin{center}
\includegraphics[width=10cm,angle=-90]{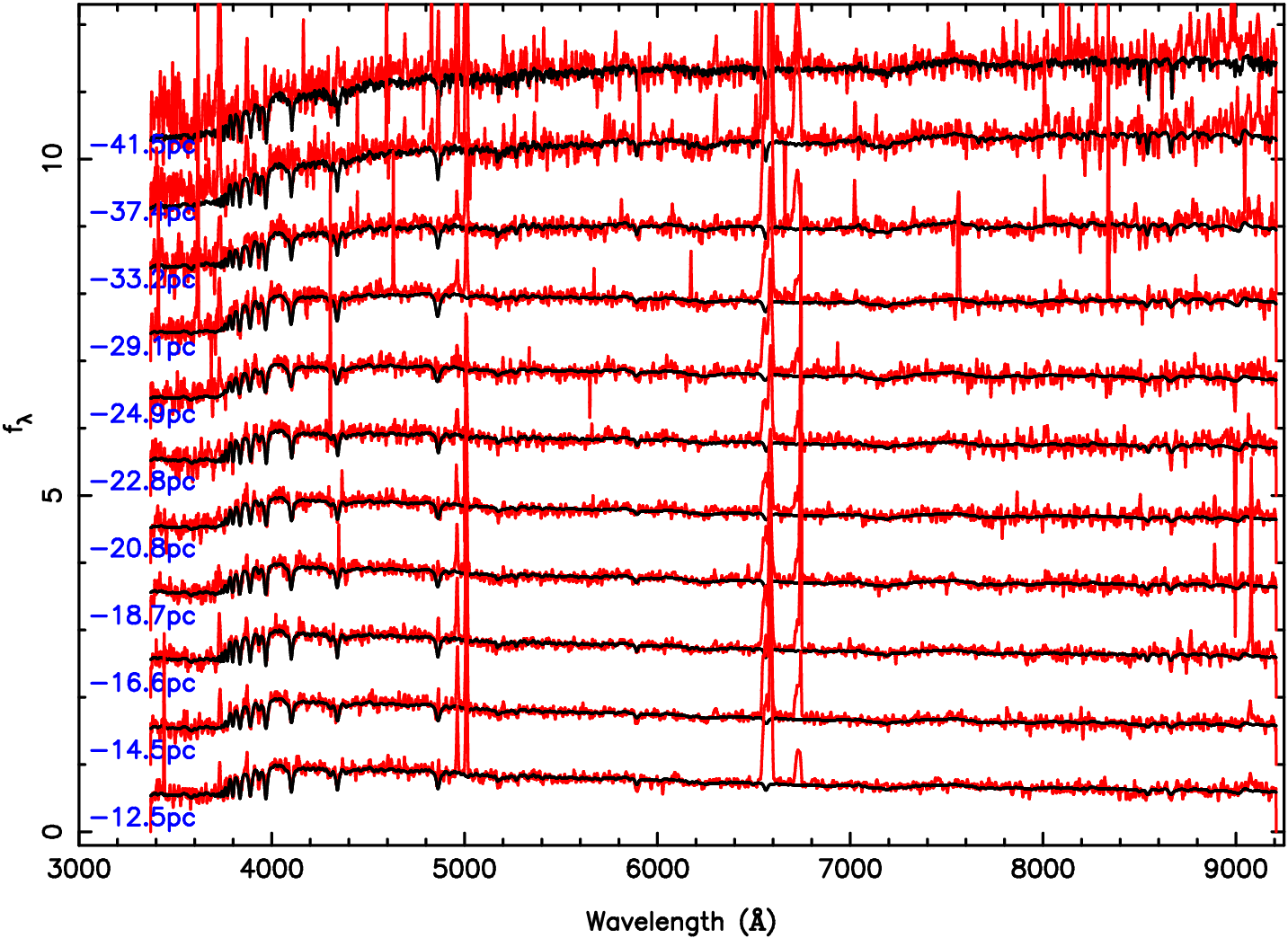}
\includegraphics[width=10cm,angle=-90]{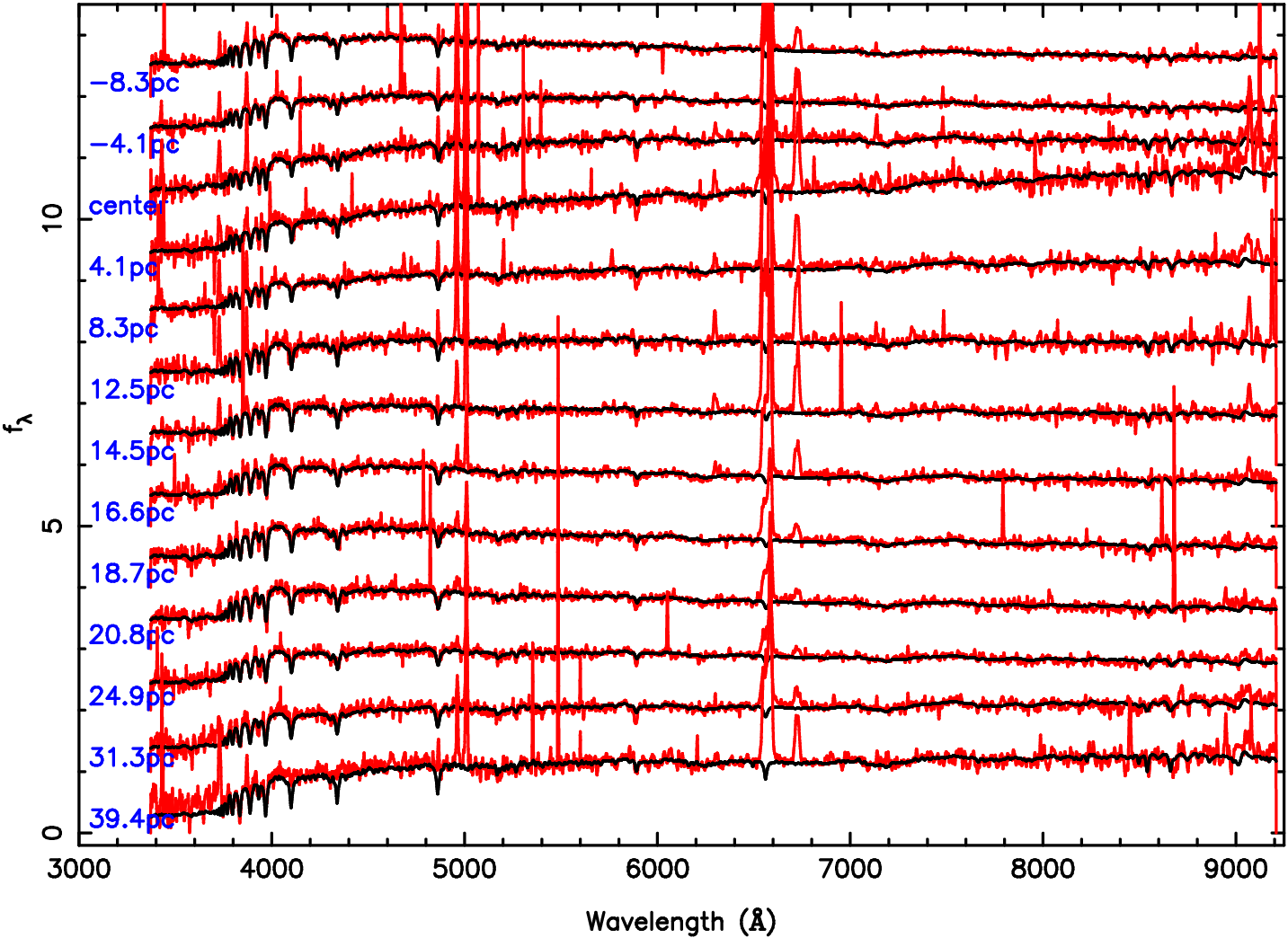}
\caption{All of 24 radial spectra with their SSP models, from -41.5
pc to 39.4 pc. The red lines are the observed spectra and the black
lines are the model spectra. The distance from the nucleus is
presented at the left of each spectrum.}
\end{center}
\end{figure*}

\begin{figure*}
\begin{center}
\includegraphics[width=10cm,angle=-90]{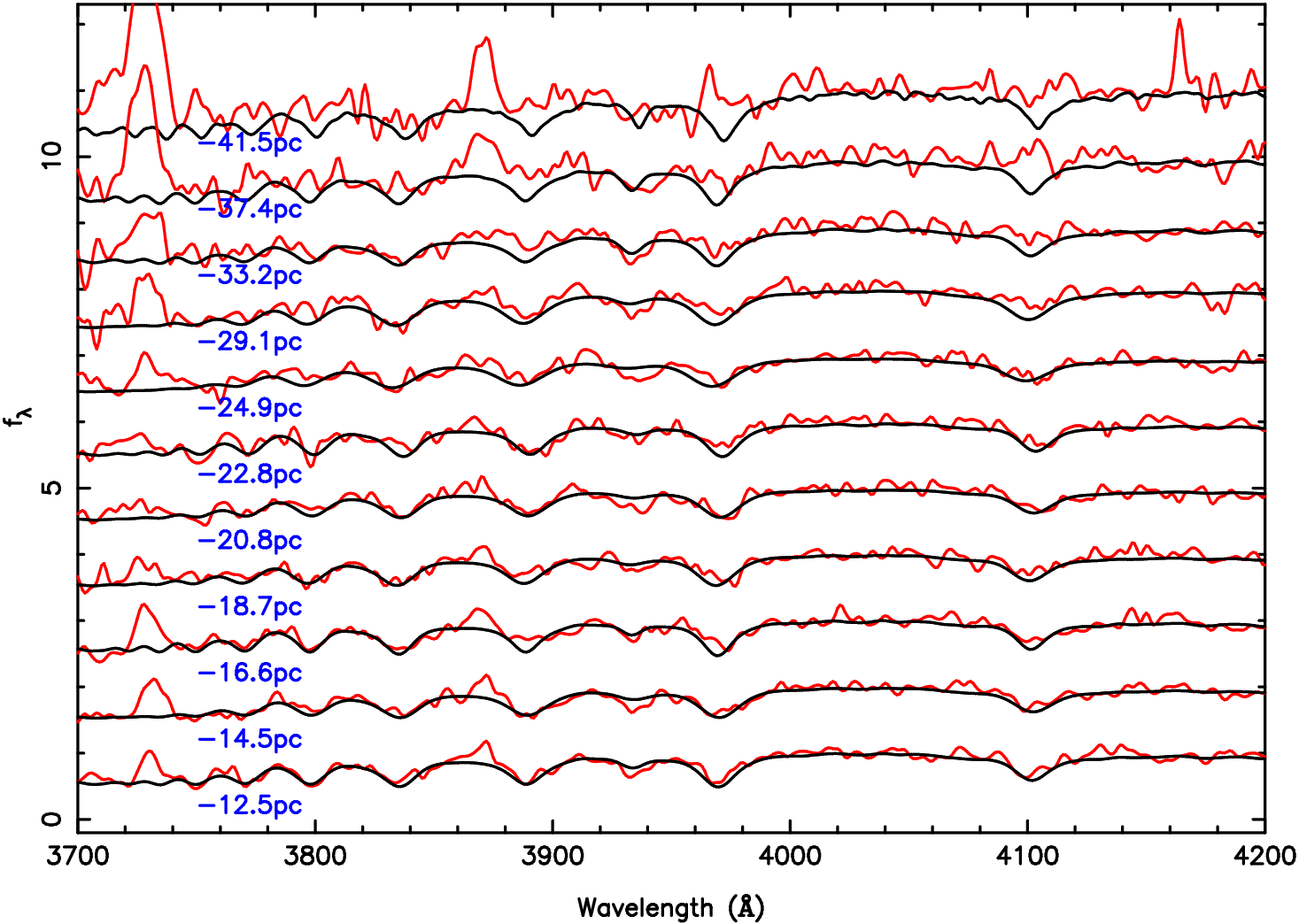}
\includegraphics[width=10cm,angle=-90]{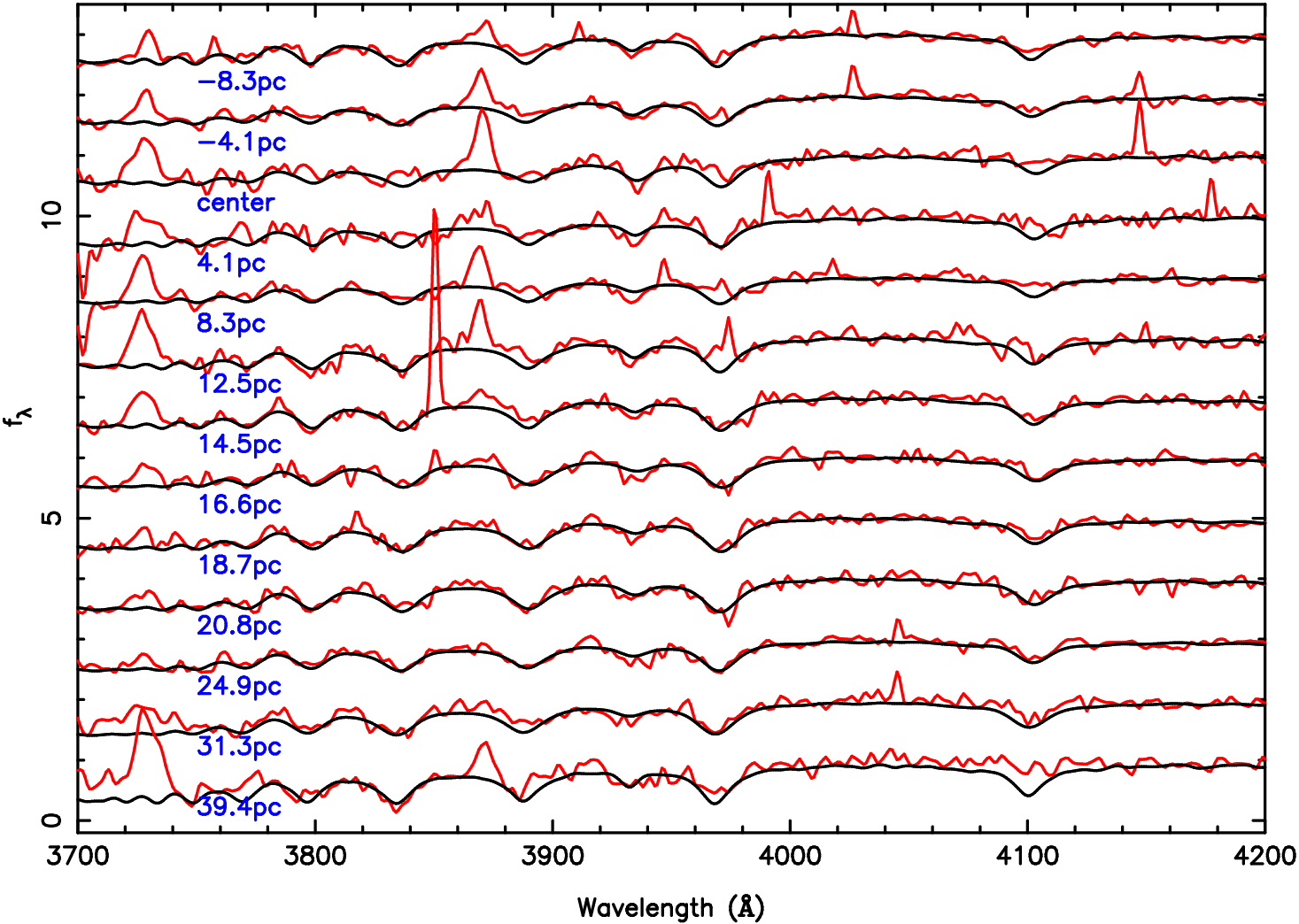}
\caption{Same as Fig. 2 but for the spectral region of
3700-4200\AA.}
\end{center}
\end{figure*}

\begin{figure*}
\begin{center}
\includegraphics[width=10cm,angle=-90]{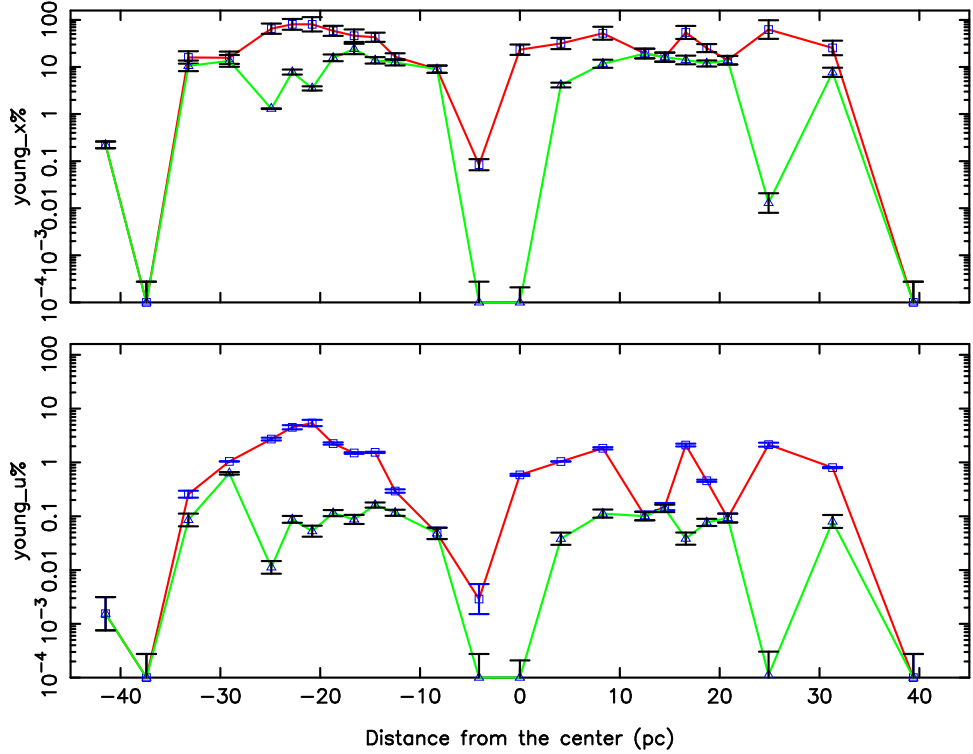}
\caption{Radial distributions of the flux fraction (top) and the
mass fraction (bottom) of the young stellar population for
$t<~0.102~Gyr$ (the red line) and for $t<~24.5~Myr$ (cyan line).
The points on the x-axis give locations where no young stellar
population is required in SSP model.}
\end{center}
\end{figure*}

\subsection{The flux/mass fraction for young stellar populations}
Through above spectral synthesis, we can obtain flux-fraction $x_j$ and mass-fraction $\mu_j$ for different stellar populations of ages and
metallicities. The mass--flux ratio can be found in the STARLIGHT
manual on the webpage, http://astro.ufsc.br/starlight/. A SSP model
for a spectrum taken from a point away from the nucleus by -12.5 pc is shown in Fig. 1.
The red curve is the observed \HST STIS spectrum by G430L and G750L.
On top left panel in Fig.1, the black curve is the SSP fit with
linear combination of different stellar population templates. In the
bottom left panel, the cyan curve is the residuals. The right panel
shows the distribution of flux fraction as function of the age of stellar
population plates (top), and the corresponding distribution of mass fraction (bottom). SSP models for all of the 24 radial spectra
are presented in Fig.2. Fig. 3 is the same as Fig. 2 but for the
spectral region of 3700-4200\AA in detail.

Properties of 24 radial spectra of M51 are presented in Table 2.
These properties are described as follows. Line 1 (top line):
the sequence number of radial spectrum; line 2: the distance
from the nucleus in the unit of pc; line 3: the size of extraction box
in pixels; line 4: the snr of radial spectrum; line 5: reduced $\chi^2$
of SSP fit; line 6: the intrinsic extinction; line 7: the flux
fraction of power law component; line 8: the flux fraction of young
stellar population ($t < ~0.102~ Gyr$); line 9: the flux fraction of
young stellar population ($t < 24.5 Myr$); line 10: the mass
fraction of young stellar population ($t < ~0.102~ Gyr$); line 11:
the mass fraction of young stellar population ($t < 24.5~ Myr$).

In Table 2, the intrinsic extinction is needed in the fit, from
0.37-1.46. Regarding the age-metallicity degeneracy, \cite{Cid05} 
mentioned that this problem is presented in STARLIGHT and
introduces systematic biases in age and Z estimates of up to
0.1-0.2 dex. From results of the simulation by \cite{Cid05}, it
is suggested that the uncertainty in SSP results can be given by the
effective starlight snr at 4020 \AA\ in three age bins, i.e.,
"young" ($t < 0.1~ Gyr$), "intermediate" ($0.1Gyr\le t \le 1Gyr$),
and "old" ($t > 1~Gyr$). For the "young" age bin, the errors in
flux/mass fraction are much smaller than those in the ages of
"intermediate" and "old". For our spectra, the effective starlight
snr at 4020 \AA\ is about 5-15. Based on table 1 in \cite{Cid05}, 
considering the "intermediate" and "old" ages, an
uncertainty of 8-14\% can be given for the flux-fraction and 7-10\% for the
mass-fraction. However, considering the "young" age, it is 1-2\% and
4-7\% for the flux-fraction, the mass-fraction, respectively. The
uncertainty in the flux/mass fraction for the age of "young" is
smaller than that for the ages of "intermediate" and "old" .
Therefore, 10\% is adopted as their typical uncertainty. In our SSP
fit, the power-law slope is fixed to -0.5. We tried other slopes,
and found the result shows little change.

In Figure 4, the flux/mass fraction is shown as a function of distance from
the nucleus of M51. It is found that the mean flux fraction of young
stellar populations ($t < 24.5~ Myr$, the cyan line in Fig. 4) is
about 9\%, excluding some regions near the center (from -6 pc to 2
pc). Excluding these regions with a flux fraction from the young stellar population of zero (from -6 pc to 2 pc), the mean mass fraction is about 0.09 \%. If choosing $t
< 0.102~ Gyr$, the mean flux fraction of young stellar populations
is 34\%, the mean mass fraction is 1.3\%, and the mass fraction is
also smaller in the central region (the red line in Fig. 4).

\subsection{Specific star formation rate}

Two widely employed methods to measure the current star formation
rates (SFR) is by examining the ultraviolet continuum, and by using the
emission lines of \ha, \oii, etc. \citep{Kennicutt98}. For a sample
of 82302 star forming galaxies from SDSS, \cite{Asari07}
investigated the current SFR derived from the SSP synthesis and
found that it is consistent with that from the H$\alpha$ SFR
indicator. They found that the Spearman rank correlation coefficients applied to results show there is a strong correlation when the critical age is selected from
10 Myr to 0.1 Gyr. In our paper, the current specific SFR (SSFR) is
defined by the mean SSFR over the past 24.5 Myr, i.e., the mass
fraction of first 12 of the 45 templates described in section 2.3
with age less than 24.5 Myr \citep{Asari07},
\begin{equation}
{\rm SSFR(t < 24.5 Myr)}=\frac{\rm
SFR}{M_*}=\frac{\sum^{12}_{j=1}\mu_j}{\rm 24.5Myr}
\end{equation}
where $M_*$ is the stellar mass.

The mean young mass fraction is 0.09\% ($t < 24.5 Myr$) and the current
$\rm SSFR(t < 24.5 Myr)$ is about $0.037\pm 0.004$ $\rm Gyr^{-1}$
(Table 2, Fig. 4). The young stellar populations are not required in
the center inner $\sim$ 8 pc in M51 (from -6.0 pc to 2.0 pc),
suggesting a possible SSFR suppression in the circumnuclear
regions from the feedback of the AGNs. Considering
the young flux/mass fraction for $t < 0.102~ Gyr$ in stead of $t <
24.5~ Myr$, the mean young mass fraction is 1.3 \% and the current
$\rm SSFR(t < 0.102~ Gyr)$ is about $0.13\pm 0.013$ $\rm Gyr^{-1}$.
The central SSFR is also smaller in the central region. We also
found that there are dips in the SSFR at the edge of our measured region, around 40 pc away from the center.

The nucleus of M51 is selected between cloud 4 and cloud 5 following
\cite{Bradley04}. Cloud 4 and cloud 5 are all at $\sim$7 pc
from the nucleus. In Fig. 14 of \cite{Bradley04}, they found
that the radio jet (8.4GHz) is elongated with a position angle
close to that of this measurement of M51 with the \HST STIS slit. From Table 2 and Fig.
3, the region where young stellar populations are not required ($t <
24.5~ Myr$) is not symmetrical with respect to the nucleus along the STIS long
slit. This suggests that the radial distribution of SSFR in M51 is
not symmetrical with respect to the direction along the long slit. The influence
of AGN feedback on the circumnuclear region is not symmetrical. This
unsymmetrical SSFR distribution is possibly due to the AGN jet in
M51 \citep[e.g.][]{Bradley04}. Using the Submillimeter Array (SMA),
which has a spatial resolution of $\sim$ 10 pc, \cite{Matsushita13}  found two
highly disturbed CO(2-1) molecular gas features, which are
perpendicular to the radio jet. These observations suggest the possible AGN
feedback from the radio jet in M51. An AGN with strong jet activity can
have less star formation in the circumnuclear regions, which is
consistent with our result. In the future, integrated field unit
(IFU) onboard the James Webb Space Telescope is needed to investigate the
AGN-starburst connection in the circumnuclear region.

\section{conclusions}
With the high spatial resolution provided by \HST STIS ($\sim$ 2pc), the radial
stellar population in the inner $\sim$ 40 pc for the nearby Seyfert 2
galaxy M51 is investigated. The main conclusions can be summarized
as follows: (1) We present 24 radial optical spectra taken with the
STIS long slit. These spectra cover the region extending $\sim
1\arcsec$ from the nucleus(i.e., -41.5 pc to 39.4 pc). (2) By SSP,
the stellar contribution to these radial optical spectra is
modeled. It is found that the mean flux fraction of young stellar
populations (younger than 24.5 Myr) is about 9\%. Excluding some
positions near the center (from -6.0 to 2.0 pc), the mean mass
fraction is about 0.09\% and the mean SSFR is about 0.037 $\rm
Gyr^{-1}$. (3) These young stellar populations are not required to be in the
center inner $\sim$ 8 pc in M51, suggesting possible SFR
suppression in the circumnuclear regions from the AGN feedback. (4)
The radial distribution of SSFR in M51 is not symmetrical with respect to the STIS long slit, which implies that the influence of
AGN feedback on the circumnuclesr region is possibly not symmetrical. This
unsymmetrical SSFR distribution is possibly due to the AGN jet in
M51.

\section{ACKNOWLEDGMENTS}
We are grateful to the anonymous referee for instructive comments.
This work has been supported by the National Science Foundations of
China (Grant Nos. 11373024; 11233003; 11173016).

\end{document}